\documentclass[twocolumn,superscriptaddress,showpacs,preprintnumbers,amsmath,amssymb]{revtex4}
\usepackage{amsmath}
\usepackage{amssymb}
\usepackage{dcolumn}
\usepackage{graphicx}
\usepackage{bm}
\usepackage{epstopdf}
\usepackage{threeparttable}
\usepackage{float}
\usepackage{color}

\begin{document}
\title{Beyond packing of hard spheres: The effects of core softness, non-additivity, intermediate-range repulsion, and many-body interactions on the glass-forming ability of  bulk metallic glasses}
\author{Kai Zhang\footnote{These authors contributed equally to this work.}} 
\affiliation{Department of Mechanical Engineering and Materials Science, Yale University, New Haven, Connecticut, 06520, USA}
\affiliation{Center for Research on Interface Structures and Phenomena, Yale University, New Haven, Connecticut, 06520, USA}

\author{Meng Fan$^*$} 
\affiliation{Department of Mechanical Engineering and Materials Science, Yale University, New Haven, Connecticut, 06520, USA}
\affiliation{Center for Research on Interface Structures and Phenomena, Yale University, New Haven, Connecticut, 06520, USA}

\author{Yanhui Liu} 
\affiliation{Department of Mechanical Engineering and Materials Science, Yale University, New Haven, Connecticut, 06520, USA}
\affiliation{Center for Research on Interface Structures and Phenomena, Yale University, New Haven, Connecticut, 06520, USA}
\author{Jan Schroers}
\affiliation{Department of Mechanical Engineering and Materials Science, Yale University, New Haven, Connecticut, 06520, USA}
\affiliation{Center for Research on Interface Structures and Phenomena, Yale University, New Haven, Connecticut, 06520, USA}
\author{Mark D. Shattuck}
\affiliation{Department of Physics and Benjamin Levich Institute, The City College of the City University of New York, New York, New York, 10031, USA}
\affiliation{Department of Mechanical Engineering and Materials Science, Yale University, New Haven, Connecticut, 06520, USA}
\author{Corey S. O'Hern}
\affiliation{Department of Mechanical Engineering and Materials Science, Yale University, New Haven, Connecticut, 06520, USA}
\affiliation{Center for Research on Interface Structures and Phenomena, Yale University, New Haven, Connecticut, 06520, USA}
\affiliation{Department of Physics, Yale University, New Haven, Connecticut, 06520, USA}
\affiliation{Department of Applied Physics, Yale University, New Haven, Connecticut, 06520, USA}

\date{\today}

\begin{abstract}
When a liquid is cooled well below its melting temperature at a rate
that exceeds the critical cooling rate $R_c$, the crystalline state is
bypassed and a metastable, amorphous glassy state forms instead. $R_c$
(or the corresponding critical casting thickness $d_c$) characterizes
the glass-forming ability (GFA) of each material. While silica is an
excellent glass-former with small $R_c<10^{-2}$~K/s, pure metals and
most alloys are typically poor glass-formers with large
$R_c>10^{10}$~K/s. Only in the past thirty years have bulk metallic
glasses (BMGs) been identified with $R_c$ approaching that for
silica. Recent simulations have shown that simple, hard-sphere models
are able to identify the atomic size ratio and number fraction regime
where BMGs exist with critical cooling rates more than $13$ orders of
magnitude smaller than those for pure metals.  However, there are a
number of other features of interatomic potentials beyond hard-core
interactions.  How do these other features affect the glass-forming
ability of BMGs?  In this manuscript, we perform molecular dynamics
simulations to determine how variations in the softness and
non-additivity of the repulsive core and form of the interatomic pair
potential at intermediate distances affect the GFA of binary 
alloys.  These variations in the interatomic pair potential allow us
to introduce geometric frustration and change the crystal phases that
compete with glass formation. We also investigate the effect of tuning
the strength of the many-body interactions from zero to the full
embedded atom model on the GFA for pure metals. We then employ the
full embedded atom model for binary BMGs and show that hard-core
interactions play the dominant role in setting the GFA of
alloys, while other features of the interatomic potential only
change the GFA by one to two orders of magnitude.  Despite their
perturbative effect, understanding the detailed form of the
intermetallic potential is important for designing BMGs with ${\rm
  cm}$ or greater casting thickness.
\end{abstract}

 \maketitle


\section{Introduction} 
\label{intro}

When metallic liquids are cooled at rates $R$ exceeding the critical
cooling rate $R_c$, crystallization can be bypassed and amorphous
alloys are formed~\cite{telford:2004}. Pure metals and most alloys are
extremely poor glass formers with $R_c > 10^{10}$ K/s. In contrast, a
number of bulk metallic glasses (BMGs) have been identified with $R_c
< 1$ K/s and critical casting thickness $d_c > 1$ cm, which enables them to be
employed in commercial applications~\cite{schroers:2013,zhong:2014}.
The discovery of novel BMGs with optimized casting thickness and
mechanical properties has largely been a trial-and-error
process~\cite{greer:1995,suryanarayana:2011}. Although combinatorial
deposition and characterization
techniques~\cite{flores:2014,ding:2014} now allow efficient
exploration of parameter space, there are an exponentially large
number of possible BMG-forming atomic
compositions~\cite{zhang:2015b}. Thus, a quantitative and predictive
understanding of the GFA of BMG-forming alloys is necessary to narrow
down the vast parameter space.

\begin{figure}
\begin{center}
\includegraphics[width=1.05\columnwidth]{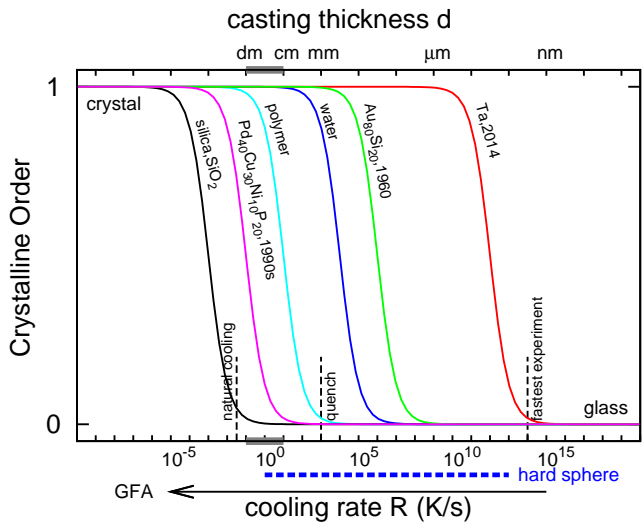}
\caption{Schematic diagram of crystalline order (such as bond
orientational order~\cite{steinhardt:1983}) versus the cooling rate $R$
in {\rm K/s} for several materials.  The critical cooling rate $R_c$
at which there is a rapid rise in the crystalline order is inversely
correlated with the material's critical casting thickness $d_c$. Smaller $R_c$
(and larger $d_c$) indicate enhanced glass-forming ability (GFA).
Pure metals, {\it e.g.} Ta, are extremely poor glass
formers~\cite{zhong:2014}. The GFA of the first fabricated metallic
alloy ${\rm Au_{80}Si_{20}}$~\cite{klement:1960} is similar to that
of water~\cite{debenedetti:2003}, but is a poor glass-former
compared to polymers~\cite{schroers:2014} and
silica~\cite{weinberg:1989}. The best bulk metallic glasses (BMGs),
{\it e.g.}  ${\rm
Pd_{40}Cu_{30}Ni_{10}P_{20}}$~\cite{yokoyama:1999}, possess {\rm
cm} or greater critical casting thicknesses and $<1$~{\rm K/s} critical
cooling rates (solid gray bars).  In recent simulations, we have
shown that hard-core atomic interactions can account for more than
$13$ orders of magnitude variation in $R_c$ (thick dashed line) from
$10^{0}$~{\rm K/s} for typical BMGs to $10^{13}$~{\rm K/s} for pure
metals~\cite{zhang:2014}.}
\label{fig:gfa}
\end{center}
\end{figure}

Silica and polymers possess critical cooling rates that are more than
$15$ and $10$ orders of magnitude lower, respectively, than those for
pure metals (Fig.~\ref{fig:gfa}). Network bonding in silica and chain
entanglement in polymers provide the physical mechanisms to inhibit
crystallization~\cite{zachariasen:1932,johnson:1999,hoy:2012}.  In
contrast, the main source of geometric frustration in alloys
is the mismatch between atomic sizes~\cite{egami:1984,miracle:2003,miracle:2004,miracle:2013,jalali:2004,jalali:2005}.
Molecular dynamics simulations of binary hard spheres have shown that
tuning the atomic size ratio can decrease $R_c$ by more than $13$
orders of magnitude~\cite{zhang:2014}.  Packing of hard spheres can
also rationalize the correlation between the number of components, their 
atomic size ratios, and
the GFA of BMGs~\cite{zhang:2015b}.

Although the packing of hard spheres plays an important role in
determining the GFA of alloys, it is obvious that metals possess
additional features that are not represented by hard-sphere
interactions. Other features of metallic interactions, such as
metallic bonding~\cite{zhang:2015}, the form of the interatomic 
pair potential, and many-body
interactions~\cite{daw:1983}, can change the crystalline structure
that competes with glass formation and change the prediction of $R_c$
by several orders of magnitude from the hard-sphere value. Compared to
the $\sim 13$ orders of magnitude variation in $R_c$ that results from
the packing of hard-spheres, changes to $R_c$ are small, but not
negligible and may explain the crucial differences between an
amorphous film and a bulk metallic glass. Since the critical casting thickness
$d_c$ is negatively correlated with $R_c$ and increasing $R_c$ by two
orders of magnitude can reduce $d_c$ by one order of
magnitude~\cite{inoue:2000}, more accurate models of intermetallic potentials
are needed to identify BMGs with $d_c > 1~{\rm cm}$
(Fig.~\ref{fig:gfa}).

\begin{figure*}
\includegraphics[width=0.68 \columnwidth]{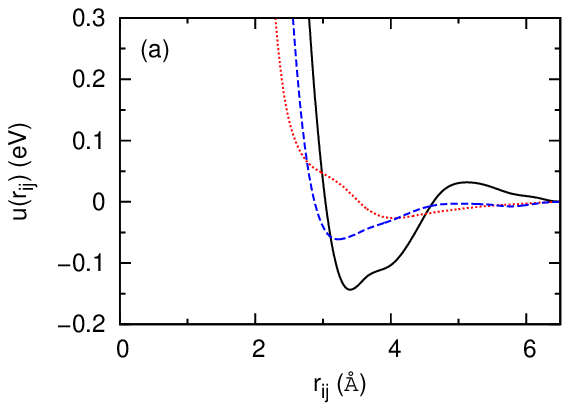}
\includegraphics[width=0.68 \columnwidth]{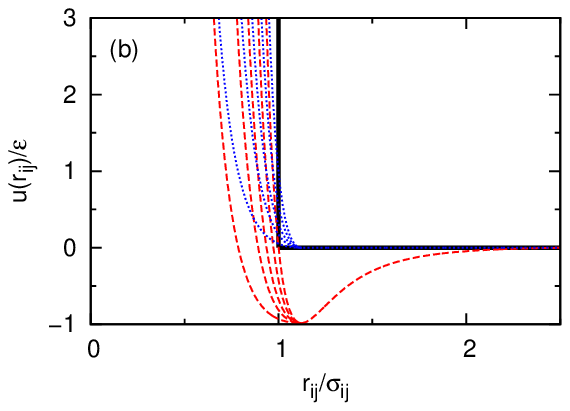}
\includegraphics[width=0.68 \columnwidth]{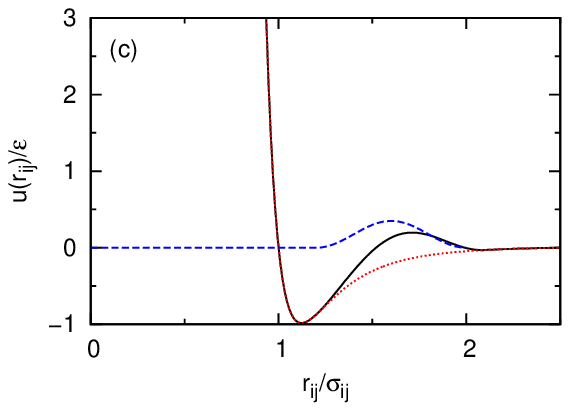}
\caption{(a) The pairwise potentials $u(r_{ij})$ (in eV) as a function of
interatomic separation $r_{ij}$ for Zr-Zr (solid line), Cu-Cu (dotted
line), and Zr-Cu (dashed line) interactions for the embedded atom
model for Zr-Cu alloys~\cite{sheng:2011}.  (b) Generalized Lennard-Jones 
(Eq.~\ref{eq:ulj})
(dashed lines) and repulsive Lennard-Jones (Eq.~\ref{eq:urlj}) (dotted lines)
interatomic potentials for several values of the core softness
exponent $m=1$, $3$, $5$, $8$, and $12$ (from left to right)
compared to the hard-sphere potential (thick solid line).  (c)
Dzugutov-Shi interatomic potential (Eq.~\ref{eq:uds}) (solid line) 
decomposed into 
the Lennard-Jones (dotted line) and sinusoidal ``bump'' potentials 
(dashed line).}
\label{fig:potential}
\end{figure*}

The interatomic potential in the embedded atom model (EAM) is
frequently implemented in computational studies of the structural and
mechanical properties, as well as the dynamics, of metallic
systems~\cite{daw:1983}.  The EAM potential energy includes a
pairwise-additive term, which is in general different from the
hard-sphere and Lennard-Jones pair potentials (Fig.~\ref{fig:potential} (a)), and a many-body
contribution from the electron charge density, which is fitted to {\it
  ab~initio} calculations of lattice parameters, elastic constants,
and other thermodynamic properties~\cite{mendelev:2007,sheng:2011}.

In this manuscript, we seek to identify the key features of the
pairwise and many-body interactions that strongly influence the GFA of
alloys.  For example, we investigate the effects of the softness
of the pairwise repulsive core, pairwise non-additivity, and the form of the
pairwise intermediate-range repulsion on the GFA. We then measure the GFA
for the full embedded atom models of several pure metals and BMGs to 
determine the contribution of the many-body interactions on the GFA. 
We find that the changes in the GFA arising from variations in the pair 
and many-body contributions of the embedded atom model are small 
compared to the $13$ orders of magnitude change in GFA between monoatomic  
and binary and ternary hard-sphere systems. However, these peturbations 
to the GFA may still be important for discovering new {\it bulk} metallic 
glass formers. 

The manuscript includes three additional sections after the
introduction.  First, in Sec.~\ref{sec:method}, we describe the
hard-sphere, repulsive Lennard-Jones, Lennard-Jones, and Dzugutov-Shi
potentials used to vary the form and non-additivity of the pairwise
interactions.  We also introduce the embedded atom model for pure
metals and alloys. For each interatomic potential, we discuss the
methods employed to measure the critical cooling rate $R_c$. We then
report the results for the GFA for all interaction potentials in
Sec.~\ref{sec:result}. We conclude the manuscript in
Sec.~\ref{sec:conclusion}. 

\section{models and methods}
\label{sec:method}

As described above, the embedded atom model for metallic systems
includes pairwise and many-body interactions. In this section, we
define three metrics (core softness, non-additivity, and intermediate-range
repulsion) to characterize the form of the pairwise interactions.  We
describe molecular dynamics simulations of monodipserse and binary
systems interacting via generalized Lennard-Jones or
Dzugutov-Shi~\cite{dzugutov:1992,shi:2014} potentials to quantify the
effects of the softness of the repulsive core and strength of the
intermediate-range repulsion on the GFA.  We also introduce molecular
dynamics simulations of binary hard spheres to study variations in the
GFA from non-additive pairwise interactions.  We estimate values for
the pairwise core softness, non-additivity, and form of the
intermediate-range repulsive interactions from fits to the pairwise
contributions of the EAM for pure metals and binary BMGs.  We also
introduce the Lennard-Jones and full EAM potentials that we employ to
study the effects of many-body interactions on the GFA.
 
\subsection{Lennard-Jones (LJ) and Repulsive Lennard-Jones (RLJ) Potentials}
\label{lj_section}

To tune the softness of the pairwise repulsive core~\cite{zhen:1983},
we employ the generalized $m$-$n$ Lennard-Jones (LJ) potential
(Fig.~\ref{fig:potential} (b)),
\begin{widetext}
\begin{equation}
\label{eq:ulj}
u_{\rm LJ}^{m-n}(r_{ij}) =
\left\{
\begin{array}{ccc}
 \epsilon \left[ 2^{\frac{m}{6}} \frac{n}{m-n} \left(\frac{\sigma_{ij}}{r_{ij}}\right)^m  - 2^{\frac{n}{6}} \frac{m}{m-n} \left(\frac{\sigma_{ij}}{r_{ij}}\right)^n\right] & , &  r_{ij} \le r_m\\
u_{\rm LJ}& , &r_{ij}  >  r_m,
\end{array}
\right.
\end{equation}
\end{widetext}
where $\sigma_{ij}=(\sigma_{i}+\sigma_{j})/2$, $\sigma_i$ is the
diameter of atom $i$, and $\epsilon$ is the energy scale of the
interaction.  The interaction potential has a minimum $u_m =
-\epsilon$ at $r_m = 2^{1/6}\sigma_{ij}$. The exponent $m$ (or equivalently 
the curvature $\kappa$ of the pair potential at the minimum) controls
the softness of the repulsive core, where smaller $m$ corresponds to
softer interactions. Note that the generalized Lennard-Jones potential
is fixed at $u_{\rm LJ}(r_{ij}) \equiv u_{\rm LJ}^{ 12-6}(r_{ij})$ for
$r_{ij} > r_m$. To separate the effects of the attractive interactions
from the repulsive core, we also studied the generalized $m$-$n$
repulsive Lennard-Jones (RLJ) potential~\cite{weeks:1971} as shown in
Fig.~\ref{fig:potential} (b):
\begin{equation}
\label{eq:urlj}
u_{\rm RLJ}^{ m-n}(r_{ij})=
\left\{
\begin{array}{ccc}
u_{\rm LJ}^{ m-n}(r_{ij}) + \epsilon & , &  r_{ij} \le r_m\\
0& , &r_{ij}  >  r_m.
\end{array}
\right.
\end{equation}

To obtain physical values for the softness exponent $m$, we fit the
repulsive part of the EAM pair potential of typical BMG-forming
elements to $u_{\rm RLJ}^{m-6}(r)$.  As shown in
Table~\ref{table:softness}, we find that
$m$ varies from approximately $3$ to $14$.  The repulsive cores for
most metals appear softer than Lennard-Jones interactions with $m=12$.

\begin{table}
\begin{center}
\begin{threeparttable}
\caption{Softness exponent $m$ from the repulsive Lennard-Jones
potential (Eq.~\ref{eq:urlj}) for the self-part of the pair
potential contribution to the embedded atom model for common atomic
species found in BMGs~\cite{sheng:2011}. The exponent $m$ varies
linearly with the curvature $\kappa$ (given in units of $\epsilon/\sigma_A^2$) of the interatomic potential at
its minimum $r_m$. 
References for the EAM potentials are provided in
columns $4$ and $8$. Several atom types possess multiple EAM potentials.}
\begin{tabular}{c c c c  c c c c}
\hline
atom   & $\kappa$ &    $m$ & Ref. &  atom & $\kappa$  &  $m$  & Ref.\\
\hline
Zr &38.77 & 8.14 & \cite{sheng:2011} & Pb &35.30  & 7.41  & \cite{zhou:2004} \\
 &31.77  &  6.67 & \cite{zhou:2004} & Mg &42.69  & 8.97  & \cite{zhou:2004}\\
 &66.91 & 14.05 & \cite{mendelev:2007} & Fe &31.41   & 6.60   & \cite{zhou:2004}\\
Ag&48.77  &10.24 & \cite{sheng:2011} & Co &32.88   & 6.90  & \cite{zhou:2004}\\
 &35.49 &7.45 & \cite{zhou:2004} &  &38.06   &7.99  & \cite{pun:2012}\\
 &40.35 & 8.47 & \cite{williams:2006} & Ta &21.36   & 4.49  & \cite{zhou:2004}\\
Al &33.48 & 7.03 & \cite{sheng:2011} &  &15.64 &3.28  & \cite{ravelo:2013}\\
 &20.65  &4.34  & \cite{zhou:2004} & &24.56  &5.16  & \cite{youhong:2003}\\
 &10.89 & 2.29 & \cite{winey:2009}  &Cu &28.66  & 6.02  & \cite{zhou:2004}\\
Ni &34.36  & 7.21 & \cite{sheng:2011} & Au &38.40  &8.06   & \cite{zhou:2004}\\
 &30.66  & 6.44 & \cite{zhou:2004} &  &48.20 &10.12   & \cite{gregory:2005}\\
 &47.96  & 10.07 & \cite{mishin:1999} & Ti &32.93   & 6.92   & \cite{zhou:2004}\\
Pd &43.72  & 9.18  & \cite{sheng:2011} & Mo &20.42   & 4.29  & \cite{zhou:2004}\\
 &33.39  & 7.01 & \cite{zhou:2004} & W &22.63 &4.75  & \cite{zhou:2004}\\
Pt &43.47  & 9.13  & \cite{sheng:2011}  &Nb &28.19   & 5.92  & \cite{fellinger:2010}\\
 &24.04  & 5.05  & \cite{zhou:2004}\\

\hline
\end{tabular}
\label{table:softness}
\begin{tablenotes}
\item 
\end{tablenotes}
\end{threeparttable}
\end{center}
\end{table}

To investigate the effects of softness of the pairwise repulsive core
on the GFA of metallic systems, we performed molecular dynamics (MD)
simulations of $N=1372$ spherical atoms with mass $m_0$ that interact
via the generalized Lennard-Jones and repulsive Lennard-Jones
potentials with $n=6$ and a range of $m$ values.  We studied three
binary LJ systems with softness exponents $m_A=m_B=m_{AB}=12$
(LJ12-6), $m_A=m_B=m_{AB}=5$ (LJ5-6), and $m_A=12$, $m_B=5$, and
$m_{AB}=8$ (LJ12-6/LJ5-6). We set the atomic diameter ratio to be
$\alpha = \sigma_{B}/\sigma_{A}=0.95$ and varied the number fraction
of small atoms $x_B = N_B/N$ from $0$ to $1$. Temperatures and times
are given in units of $\epsilon/k_B$ and
$\sigma_{A}\sqrt{m_0/\epsilon}$, respectively.  After equilibrating
the systems at high temperature $T_0=2$, the liquids were cooled
exponentially $T(t)=T_0 \exp (-Rt)$ with rate $R$ to low temperature,
$T_f=0.01$, using the Gaussian constraint thermostat~\cite{allen:1987}
with time step $\Delta t=0.001$. Constant volume $V$ simulations
at number density $\rho \sigma_{A}^3 =N\sigma_A^3/V=1$ were performed
for both the LJ and RLJ models. For LJ systems, we also cooled 
systems with the constraint that the pressure $p$ (in units of
$\epsilon/\sigma_A^3$) decreased exponentially in time from an initial
pressure $p_0=1$ to final pressure $p_f=0.001$ according to
\begin{equation}
\label{eq:p}
p(t)=p_0 \exp \left[-\frac{\log
    (p_0/p_f)}{\log (T_0/T_f)}Rt\right]
\end{equation}
using a Gaussian constraint
barostat~\cite{allen:1987}.  A cooling rate of $R=1$ in the units used
in the MD simulations corresponds to a cooling rate of $10^{15}$ K/s
using $\sigma_A\sim3\times10^{-10}$ m, $\epsilon/k_B\sim10^3$ K, and
molar mass $M\sim10^{-1}$ kg/mol, which are typical values for BMGs~\cite{zhen:1983}.

\subsection{Non-additive binary hard spheres}
\label{sec:na}

The sizes of metallic atoms are often estimated from the first peak of the
radial distribution function $g(r)$ of crystalline and disordered
solids~\cite{humerothery:1950}. In binary alloys with species $A$ and
$B$, the repulsive core $\sigma_{AB}$ between atoms $A$ and $B$ can
differ from the average diameter $\overline{\sigma}_{AB} =
(\sigma_{A} + \sigma_{B})/2$. We quantify the non-additivity of the
pairwise repulsive core using the parameter 
\begin{equation}
\Sigma = \frac{\sigma_{AB}}{{\overline \sigma}_{AB}} - 1.
\end{equation}
Many binary alloys possess $\Sigma<0$, which indicates that the
repulsive core $\sigma_{AB}$ between $A$ and $B$ atoms is smaller
than the average diameter.  We list $\sigma_{A}$, $\sigma_{B}$,
$\sigma_{AB}$, and $\Sigma$ for several binary alloys obtained from
EAM calculations of $g(r)$ in Table~\ref{table:nonadd}.  Non-additive
binary hard spheres have been shown to form exotic crystalline
structures, in particular intermetallic
compounds~\cite{punnathanam:2006,woodcock:2011}.  In addition,
non-additivity due to bond shortening with $\Sigma < 0$ can
lead to unusual intermediate-range order in
BMGs~\cite{cheng:2009,senkov:2012,sheng:2006}.  The well-studied
Kob-Andersen model for $\rm{Ni}_{80}{\rm P}_{20}$ glasses also has
$\Sigma=-0.149$~\cite{kob:1995}.

\begin{table}
\begin{center}
\begin{threeparttable}
\caption{Atomic diameters ($\sigma_{A}$ and $\sigma_{B}$ in  $\AA$) determined 
by the first peak of the 
radial distribution function $g(r)$ obtained from EAM simulations of 
several binary alloys~\cite{sheng:2011}. We also list $\sigma_{AB}$
from $g(r)$, the diameter ratio $\alpha$, and the non-additivity parameter $\Sigma$.} 
\begin{tabular}{c  c   c c  c c }
\hline
Alloy   &    $\sigma_{A}$  & $\sigma_{B}$ & $\sigma_{AB}$ & $\alpha$ & $\Sigma$\\
\hline
Zr-Cu &  3.15 & 2.49 & 2.75  & 0.79& -0.025 \\
Ni-P&  2.57 & 2.19 & 2.23 &0.85& -0.063 \\
Zr-Ni &  3.23 & 2.43 & 2.69 &0.75& -0.049 \\
Zr-Al &  3.21 & 2.69 & 2.93 &0.84& -0.007 \\
Ag-Al &  2.87 & 2.69 & 2.69 &0.94 &  -0.032\\
Mg-Cu &  3.11 & 2.47 & 2.69 & 0.79&  -0.05\\
Mg-Ti &  2.97 & 2.77 & 2.99 & 0.93&  0.042\\
Y-Mg &  3.51 & 3.03 & 3.27 & 0.86&  0\\
Pd-Si &  2.97 & 2.39 & 2.51 & 0.80&  -0.063\\
Zr-Pt &  3.39 & 2.91 & 2.73 & 0.86&  -0.1333\\
Cu-Ni &  2.51 & 2.45 & 2.49 & 0.98&  0.004\\
Mg-Al & 2.99 & 2.81 & 2.99 & 0.94 & 0.031\\
\hline
\end{tabular}
\label{table:nonadd}
\end{threeparttable}
\end{center}
\end{table}

To study the effects of nonadditivity on the GFA, we compressed $N=500$
binary hard spheres with mass $m_0$ that interact pairwise via
\begin{equation}
\label{eq:uhs}
u_{\rm HS}(r_{ij}) =
\left\{
\begin{array}{ccc}
\infty & , &  r_{ij}\le \sigma_{ij}\\
0& , &r_{ij}> \sigma_{ij}
\end{array}
\right.
\end{equation}
over a range of diameter ratios $\alpha$ and number fractions of the 
small sphere $x_{B}$
using event-driven MD simulations. We first equilibrated liquid states
at packing fraction $\phi=0.25$. To compress the system, we ran the MD
simulations at constant volume for a time interval $\tau$, and then
compressed the system instantaneously until the closest pair of
spheres came into contact~\cite{jalali:2004,zhang:2014}. We performed 
successive compressions until the pressure increased to $10^3$, which
corresponds to $(\phi_J-\phi)/\phi_J < 10^{-3}$, where $\phi_J$ is 
the packing fraction at the onset of jamming.  We varied the
compression rate $R\equiv 1/\tau$ over $5$ orders of
magnitude~\cite{zhang:2014}. We report $R$ in units of
$\sqrt{k_B T/m_0\sigma_A^2}$. Note that in these units $R=1$ corresponds to a 
cooling rate of $10^{12}$ K/s for alloys~\cite{truskett:2000}.

\subsection{Dzugutov-Shi (DZ) potential}
\label{dz_section}

The pair potential of many metallic systems includes intermediate-range
repulsive interactions~\cite{Friedel:1958} in addition to short-range
attractive interactions, which can give rise to intermediate-range positional
order~\cite{fujima:2007,wu:2013}. Intermediate-range pairwise repulsive
interactions are often modeled using the Dzugutov
potential~\cite{dzugutov:1992,dzugutov:1993,roth:2000,doye:2001}. Shi
{\it et.~al.} introduced a modified version of the original Dzugutov
potential that allows one to continuously tune the interaction
potential between the LJ potential to one that includes intermediate-range
repulsion~\cite{shi:2014}. The Dzugutov-Shi (DZ) potential is given by 
\begin{equation}
\label{eq:uds}
u_{\rm DZ}(r_{ij}) =
u_{\rm LJ}(r_{ij}) + u_{\rm bump}(r_{ij}),
\end{equation}
where the ``bump'' potential $u_{\rm bump}(r_{ij})$ models  
the intermediate-range repulsive interactions using a sinusoidal pulse,
\begin{equation}
\label{eq:uds2}
u_{\rm bump}(r_{ij}) =
\left\{
\begin{array}{ccc}
\xi\sin^2\left(\pi \frac{r_{ij}/\sigma_{ij}-\lambda}{\delta- \lambda}\right) & , &  \lambda \le r_{ij}/\sigma_{ij} \le \delta\\
0& , &{\rm otherwise},
\end{array}
\right.
\end{equation}
of the strength $\xi$ within the range $\lambda\sigma_{ij} \le r_{ij} \le \delta\sigma_{ij}$. The
location of the peak and width of $u_{\rm bump}$ are given by
$(\lambda+\delta)/2$ and $\delta - \lambda$. To obtain physical values
for $\xi$, $\lambda$, and $\delta$, we fit the DZ potential to the EAM
pair potential for several elements.  We show values 
of $\xi$, $\lambda$, and $\delta$ for elements commonly found 
in BMGs in Table~\ref{table:bump}.  Pb, Pd, Pt, Mg, Fe, Ta, Au, Ti, 
Mo, W, and Nb do not have significant intermediate-range repulsive 
interactions.  

\begin{table}
\begin{center}
\begin{threeparttable}
\caption{Values of the parameters $\xi$, $\lambda$, and $\delta$
(Eq.~\ref{eq:uds}) that describe the strength and range of the
Dzugutov-Shi interatomic potential fit to the self-part of the pair 
potential of the 
embedded atom model for several atomic species. The 
fifth column provides references for the EAM for each atom type.  
}
\begin{tabular}{c  c  c  c  c}
\hline
atom   &    $\xi$ (eV) &   $\lambda$  & $\delta$  & Ref.\\
\hline
Zr & 0.42 & 1.16 & 2.24 & \cite{sheng:2011}\\
Ag & 0.16 &1.29 &2.20 & \cite{sheng:2011}\\
Cu & 0.43 & 1.18 & 1.73 & \cite{sheng:2011}\\
Ni & 0.38 & 1.19 & 1.72 & \cite{sheng:2011}\\
  Al & 0.10  & 1.76  & 2.35 & \cite{sheng:2011}\\
  &0.26   & 1.29   & 1.99 & \cite{winey:2009}\\
  Co &0.12 & 1.56 & 2.66 & \cite{pun:2012}\\
\hline
\end{tabular}
\label{table:bump}
\begin{tablenotes}
\item
\end{tablenotes}
\end{threeparttable}
\end{center}
\end{table}

To study the effects of intermediate-range repulsive interactions on
the GFA, we performed MD simulations of $N=1372$ spherical atoms that
interact pairwise via the DZ potential.  We followed the same cooling
protocol as used for the simulations of Lennard-Jones systems with
pressure that decreases exponentially in time as discussed in
Sec.~\ref{lj_section}.  We fixed the strength of the intermediate-range 
repulsive interactions at $\xi=0.35\epsilon$
and varied $\lambda$ and $\delta$ to tune the location of the peak
$(\lambda+\delta)/2$ and range $\delta-\lambda$ of $u_{\rm bump}$.  We
also studied binary mixtures composed of $A$ atoms that interact via
the DZ potential with $\xi=0.35\epsilon$, $\lambda=1.2$, and $\delta=2.15$,
and $B$ atoms that interact via the LJ potential with diameter ratio
$\alpha=0.95$.  The number fraction of small atoms $x_B$ is varied from $0$ to $1$ in
steps of $0.2$.

\subsection{LJ-EAM and EAM potential}
\label{eam_section}

The total potential energy $U$ employed in the embedded-atom model for
metals includes pairwise and many-body contributions:
\begin{equation}
\label{eq:eam}
U=\sum\limits_{i<j} u(r_{ij}) + \sum\limits_{i}F_i (\overline{\rho}^e_i),
\end{equation}
where the many-body embedding function $F_i$ depends on the electron
density associated with each atom $i$ (normalized by $e/\sigma_A^3$) and
$\overline{\rho}^e_i = \sum\limits_{j\ne i}
\rho^e(r_{ij})$~\cite{daw:1983,mendelev:2007,sheng:2011}. To quantify
the effects of the many-body interactions on the GFA, we focused on
the LJ-EAM potential, where $u(r_{ij})=u_{LJ}(r_{ij})$,
$F_i(\overline{\rho}^e_i)= A\overline{\rho}^e_i (\ln
\overline{\rho}^e_i-r_m/\sigma_A)/2$ and $\rho^e(r_{ij}) =
C\exp[-\beta(r_{ij}-r_m)]$, where $C$ and $r_m$ are calibrated to
experimental data on alloys~\cite{baskes:1999,nam:2007}. We set the
atomic diameter $\sigma_A = 2.8$~${\AA}$ and attraction depth
$\epsilon=0.2$ eV for the LJ potential to match the pair potential of
typical metals such as Zr.  The parameters $A$ and $\beta$ control the
many-body interaction strength and inverse decay length of the electron
density, respectively.

We performed MD simulations of the LJ-EAM for several pure metals and of the full EAM for several 
binary alloys using the LAMMPS simulation software~\cite{plimpton:1995}.  We cooled systems
in the liquid state to low temperature at constant zero pressure at
different rates $R$.  The initial and final
temperatures for several systems (specified by $A$ and $\beta$) are
summarized in Table~\ref{table:coolingtemperature}. For our studies of
the full EAM potential, we set $N=4000$ and fixed the initial and final temperatures 
at $T_i=2000K$ and $T_f=300K$.

\begin{table}
\begin{center}
\begin{threeparttable}
\caption{The initial and final temperatures, $T_{i}$ and $T_{f}$, employed 
during the cooling protocol in the molecular dynamics simulations of the 
LJ-EAM potential with many-body interaction strength $A$ and electron 
density inverse decay length $\beta$.}
\begin{tabular}{c  c     c c     c  c     c c}
\hline
$A$ (eV) & $\beta$~(${\AA}^{-1}$)  & $T_{i} (K)$ & $T_{f} (K)$ & $A$ (eV)  &    $\beta$~(${\AA}^{-1}$) & $T_{i} (K)$ & $T_{f} (K)$\\
\hline
0 &  4 & 2000  & 300 & 0.66 &  2 & 2000  & 300 \\
1.32 &  2 & 2305  & 343 & 1.98 &  2 & 3285  & 479 \\
0.66 &  4 & 2000  & 300 & 1.32 &  4 & 2257  & 336 \\
1.98 &  4 & 3253  & 475  & 0.66 &  6 & 2000  & 300\\
1.32 &  6 & 2242  & 337 & 1.98 &  6 & 3271  & 472 \\
\hline
\end{tabular}
\label{table:coolingtemperature}
\end{threeparttable}
\end{center}
\end{table}

\subsection{Critical cooling rate}

To calculate the critical cooling rate $R_c$ for each metallic system,
we initialized the liquid state at high temperature, cooled the system
exponentially to low temperature at a given rate $R$ at either fixed
volume or exponentially decaying pressure as in Eq.~\ref{eq:p}, and
measured the global bond orientational order parameter
$Q_6$~\cite{steinhardt:1983}.  For hard-sphere interactions, we
compressed the systems so that the packing fraction approached that at
jamming onset exponentially, which is thermodynamically equivalent to
cooling systems exponentially~\cite{parisi:2010}. For all systems 
studied, the average global
bond orientational order parameter $Q_6$ versus $\log R$ possesses a
sigmoidal shape with a midpoint defined by $R_c$.  Below, we show results for
$R_c$ for the pair potentials described in
Secs.~\ref{lj_section}-\ref{dz_section} and the full and LJ-EAM potential in
Sec.~\ref{eam_section}.

\section{Results}
\label{sec:result}

\subsection{Core Softness}

\begin{figure}
\includegraphics[width=3.5in]{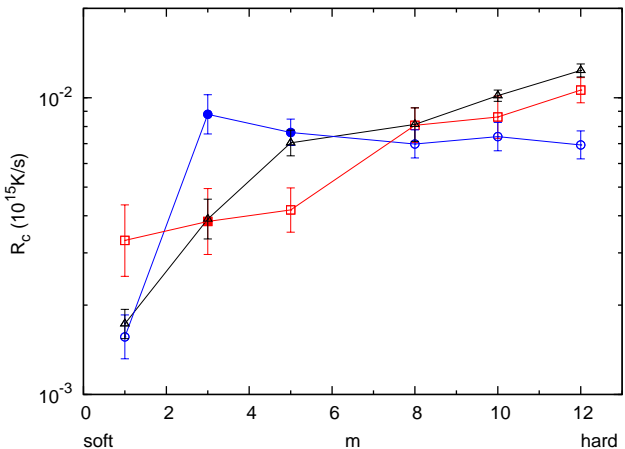}
\caption{The critical cooling rate $R_c$ (in units of $10^{15}$ {\rm
K/s}) as a function of the repulsive core softness exponent $m$ in
monodisperse systems with generalized repulsive Lennard-Jones (triangles) and
Lennard-Jones (squares) interactions cooled at constant density
$\rho \sigma_A^3=1$ and Lennard-Jones interactions with an exponentially
decaying pressure given in Eq.~\ref{eq:p} (circles). Variations in
the softness exponent lead to different crystalline structures 
that compete with glass formation including
face-centered cubic (FCC; empty symbols) and body-centered cubic
(BCC; filled symbols).}
\label{fig:LJcore}
\end{figure}

To investigate the effects of softness of the repulsive core on the
GFA, we first measured the critical cooling rate $R_c$ for
monodisperse systems that interact via the generalized LJ
(Eq.~\ref{eq:ulj}) and RLJ (Eq.~\ref{eq:urlj}) pairwise potentials as
a function of the softness exponent for $m=1$, $3$, $5$, $8$, $10$,
and $12$.  As shown in Fig.~\ref{fig:LJcore}, when cooling at constant
number density $\rho \sigma_A^3 =1$, the GFA increases weakly ($R_c$
decreases by less than an order of magnitude) as the repulsive core
becomes softer ($m$ decreases).  When cooling a LJ system with a
pressure that decays exponentially in time as in Eq.~\ref{eq:p}, the
dependence of $R_c$ on the softness exponent $m$ is even weaker,
except for systems with extremely soft core repulsions with $m=1$.
In contrast, most atomic species that are found in BMGs possess $m>4$ (Table~\ref{table:softness}). 

As shown in Fig.~\ref{fig:LJcore}, the crystalline structures that
compete with glass formation in systems with core-softened RLJ
interactions at $\rho\sigma_A^3=1$ are face-centered cubic (FCC) for
all exponents $m$ studied.  In addition, FCC
crystals compete with glass formation in LJ systems, but as the
repulsive core softens, body-centered cubic (BCC) crystals become more
stable~\cite{hoover:1972}.  We find that BCC is the crystal type that
competes with glass formation for $m=3$ LJ systems cooled at constant
density $\rho\sigma_A^3=1$ and for $m=3$ and $5$ LJ systems cooled
such that the pressure obeys Eq.~\ref{eq:p}.

\begin{figure}
\includegraphics[width=3.5in]{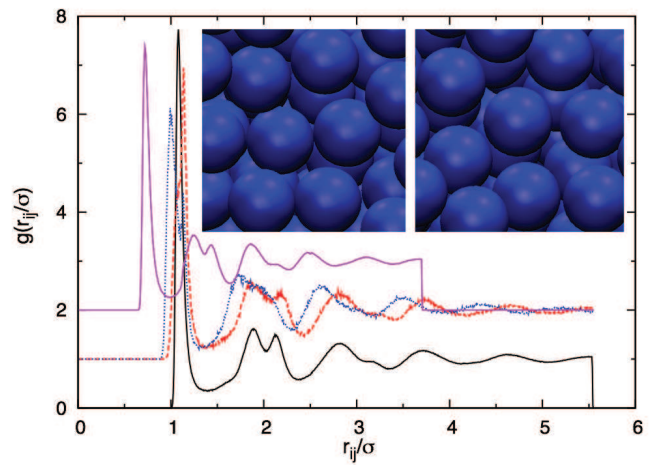}
\caption{Radial distribution functions $g(r_{ij})$ (vertically shifted
for visualization) for monodisperse spherical atoms with 
diameter $\sigma$ that interact via the
generalized $m$-$6$ LJ potential (Eq.~\ref{eq:ulj}) with $m=1$
(top), $3$ (middle), and $12$ (bottom) cooled at rate
$R=0.1>R_c$. Compared to the $m=12$ LJ system, the soft $m=1$ LJ system
shows strong volume contraction with the first peak shifted to smaller 
separations $r_{ij}$. Systems with intermediate softness $m=3$ display 
two isostructural states: contracted high density (dotted line) and
expanded low density (dashed line) glasses. The left and right insets show 
snapshots of the contracted and expanded $m=3$ LJ systems, respectively.}
\label{fig:grLJ}
\end{figure}

Structural characterizations of atomic systems that interact via the
generalized LJ potential are shown in Fig.~\ref{fig:grLJ} for cooling
rates $R > R_c$.  As the repulsive core of the potential becomes
softer ({\it i.e. $m$} decreases), the attractive well of the
potential widens to include second-neighbor attractive interactions,
which can compensate repulsive first-neighbor interactions.  Indeed,
LJ systems with $m=1$ and $3$ exhibit phase separation into dilute and
compressed regions when cooled at fixed density $\rho\sigma_A^3=1$ and
volume contraction, where the first neighbor separations are smaller
than the location of the potential minimum, when cooled such that the 
pressure obeys Eq.~\ref{eq:p}. In fact, the $m=3$ LJ system displays
two isostructural glassy states, contracted and expanded, with
different densities as shown in the inset to Fig.~\ref{fig:grLJ}.
Similar isostructural transitions have been found in equilibrium
systems with narrow-ranged attractive
interactions~\cite{frenkel:2006}. Large density differences between
polymorphs in metallic glasses such as those found in
Ce$_{55}$Al$_{45}$ are often attributed to electronic many-body
interactions~\cite{sheng:2007}. However, here we show that softening
the pairwise repulsive core (which increases the range of the
attractive well) can also give rise to polymorphs with different
densities.

\begin{figure}
\includegraphics[width=3.5in]{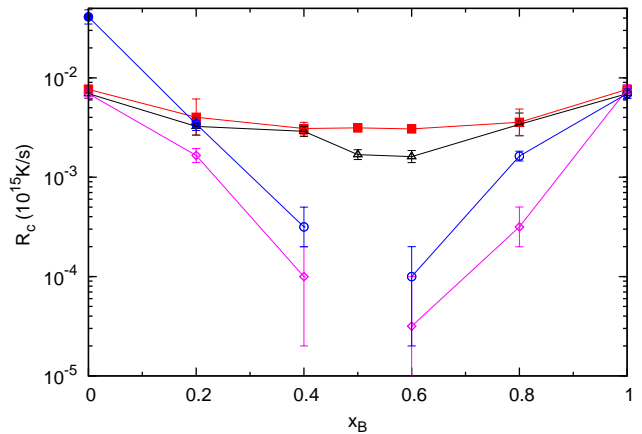}
\caption{The critical cooling rate $R_c$ for binary mixtures of spherical
atoms that interact via the generalized $m$-$6$ LJ potential (Eq.~\ref{eq:ulj})
at diameter ratio $\sigma_B/\sigma_A=0.95$ is plotted as a function of the number
fraction of small atoms $x_B$. We show $m=5$ (squares) and $12$ (triangles) 
LJ systems, mixtures (diamonds) of $m=12$ (species $A$) and $5$ 
(species $B$), as well as mixtures (circles) of spheres with DZ ($A$ species) 
and $m=12$ LJ ($B$ species) interactions. Open and filled symbols indicate
that the crystalline structure that competes with glass formation 
is FCC and BCC, respectively.}
\label{fig:rcbinary}
\end{figure}

We also investigated the effects of core softness on the glass-forming
ability in binary mixtures that interact via the generalized $m$-$6$
LJ potential. We focused on three mixtures with diameter ratio
$\alpha=\sigma_B/\sigma_A=0.95$: (1) conventional LJ systems with
$m=12$, (2) core softened LJ systems with $m=5$, and (3) mixtures of
LJ systems with $m=12$ ($A$ species) and $m=5$ ($B$ species). While
FCC is the crystalline structure that competes with glass formation
for binary LJ systems with $m=12$, BCC is the competing crystalline
structure for binary mixtures with $m=5$ for all number fractions
$x_B$ as shown in Fig.~\ref{fig:rcbinary}. For both $m=12$ and $m=5$
systems, the variation in $R_c(x_B)$, which is less than an order of
magnitude, is controlled by the diameter ratio $\alpha=0.95$. In
binary mixtures of LJ systems with $m=12$ and $m=5$ interactions, FCC
remains the crystalline structure that competes with glass formation,
except when $x_B \approx 1$.  However, because of the incompatibility
between FCC and BCC crystalline structures, the GFA for the $m=12$ and
$m=5$ LJ mixtures is significantly enhanced compared to glasses with
$m=12$ or $m=5$ interactions alone.  For example, Ni-Ta is a good
glass former despite the fact that it possesses a diameter ratio near
unity ($\alpha \approx 0.9$)~\cite{wang:2010}.  Incompatibility between competing BCC
and FCC crystal structures is a possible cause of the enhanced
GFA. As shown in
Table~\ref{table:softness}, Ni has a relatively large pairwise
repulsive exponent ($6 < m < 10$) with equilibrium FCC structure,
while Ta has a relatively small exponent ($3 < m < 5$) with equilibrium
BCC structure~\cite{humerothery:1950}. Since the softness exponents of the pairwise
interactions vary significantly from one element to another
(Table~\ref{table:softness}), softness-induced competing crystal
incompatibility can enhance the GFA of binary and multi-component
BMG-forming alloys.

\subsection{Non-additivity}
\label{non-add}

We performed event-driven molecular dynamics simulations of binary
non-additive hard spheres (Sec.~\ref{sec:na}) to investigate the
effects of non-additivity of the pairwise repulsive interactions on
the GFA of alloys.  We measured the critical cooling rate $R_c$
of non-additive binary hard spheres with diameter ratios
$\alpha=\sigma_{B}/\sigma_{A}=1.0$, $0.97$, $0.95$, $0.93$, $0.9$, and
$0.5$ and number fractions of the small spheres $x_B=0.5$ and $2/3$ over a range of
non-additivity parameters $\Sigma$.  Since $\Sigma > 0$ is rare among
binary alloys (Table~\ref{table:nonadd}), we expect that hard-sphere
systems with positive non-additivity are poor glass-formers. For
example, we find that systems with $\alpha=1$ and $\Sigma=0.05$
display strong demixing between $A$ and $B$ particles and  
are not good glass formers.

\begin{figure}
\includegraphics[width=3.5in]{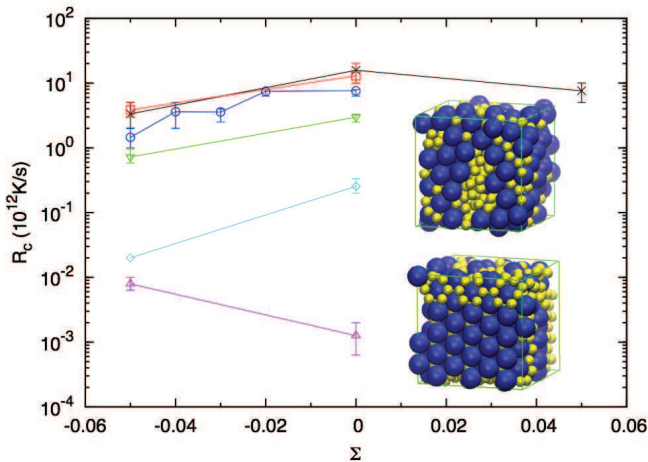}
\caption{Critical cooling rate $R_c$ (in units of $10^{12}$ K/s)
plotted as a function of the nonadditivity parameter $\Sigma$ for binary hard-sphere
systems for several diameter ratio and small-sphere number fraction combinations:
$\alpha=1.0$ and $x_B = 0.5$ (crosses), $0.97$ and $0.5$ 
(squares), $0.95$ and $0.5$ (pentagons),
$0.93$ and $0.5$ (downward triangles), $0.90$ and $0.5$
(diamonds), and $0.5$ and $2/3$ (upward triangles). The 
inset shows snapshots of (top) demixed and (b) compound
crystals that form for $R < R_c$ at $\Sigma=0.0$ and
$-0.05$, respectively, with diameter ratio $\alpha=0.5$.}
\label{fig:HSnonadd}
\end{figure}

Our previous studies of additive binary hard spheres ($\Sigma=0$) have
shown that well-mixed FCC solid solutions are the crystal structures
that compete with glass formation when $\alpha \gtrsim 0.8$, while the
systems tend to demix when $\alpha \lesssim 0.8$~\cite{zhang:2014}.  For
$\Sigma<0$ and $\alpha=1.0$, $0.97$, $0.95$, $0.93$, and $0.9$, the
GFA improves as $\Sigma$ becomes more negative, and the competing
crystal structure remains the FCC solid solution
(Fig.~\ref{fig:HSnonadd}). The change in $R_c$ with decreasing
$\Sigma$ also increases as $\alpha$ decreases with roughly an order of
magnitude difference in $R_c$ between systems with $\Sigma = 0$ and
$\Sigma=-0.05$ at $\alpha=0.9$.  Enhancement of the GFA arising from
non-additivity of the repulsive cores ($\Sigma < 0$) has also been observed in LJ systems~\cite{zhang:2013}.

For binary systems with large atomic size differences ({\it i.e.}
$\alpha \ll 0.8$), the variation of $R_c$ with $\Sigma$ is
opposite to that obtained for binary systems with small atomic size
differences. As shown in Fig.~\ref{fig:HSnonadd}, we find that $R_c$
grows with increasing $\Sigma$ at $\alpha = 0.5$. For $\alpha = 0.5$
and $\Sigma <0$, compound crystals are the ordered structures that
compete with glass formation since negative non-additivity promotes
mixing. As an example, although the $AB_2$ compound is the densest
crystal for binary hard spheres with $\alpha=0.5$ and $\Sigma=0$, it
is not kinetically accessible during compression due to the strong
drive for demixing~\cite{filion:2009,hopkins:2011,zhang:2014}.
However, when $\Sigma$ becomes negative ({\it e.g.} $\Sigma = -0.05$),
we find that the $AB_2$ compound forms easily for the compression
rates that we studied, as shown in the inset to
Fig.~\ref{fig:HSnonadd}.  Thus, the formation of intermetallic
compounds in alloys can be enhanced by pairwise negative non-additivity among
different atomic species.

\subsection{Intermediate-range Repulsive Interactions}
\label{mrr}

\begin{figure}
\includegraphics[width=3.5in]{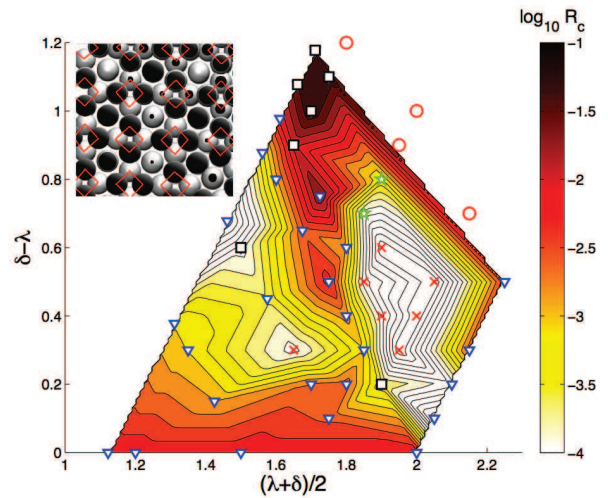}
\caption{Contour plot of the critical cooling rate $R_c$ (in reduced unit) for
monodisperse spheres that interact via the DZ potential
(Eq.~\ref{eq:uds}) as a function of the location of the peak
$(\lambda+\delta)/2$ and width $\delta-\lambda$ of $u_{\rm bump}$.
The bounds for the parameters are determined by $\lambda>r_m/\sigma_A=1.12$
and $\delta<r_c/\sigma_A=2.5$.  $R_c$ contours are interpolated from
simulation data points. The symbols indicate where FCC (triangles), BCC (squares),
$8$-$4$ (stars) crystalline structures, and microphase
separation (circles) is observed. Crosses indicate systems for
which the competing crystal structure is unknown and $R_c$ is
estimated from the slowest cooling rate employed. The inset shows a snapshot of
a $8$-$4$ crystal that includes top (dark) and bottom (light) layers
of atoms with square symmetry (red squares).}
\label{fig:RcDZ}
\end{figure}

\begin{figure}
\includegraphics[width=3.5in]{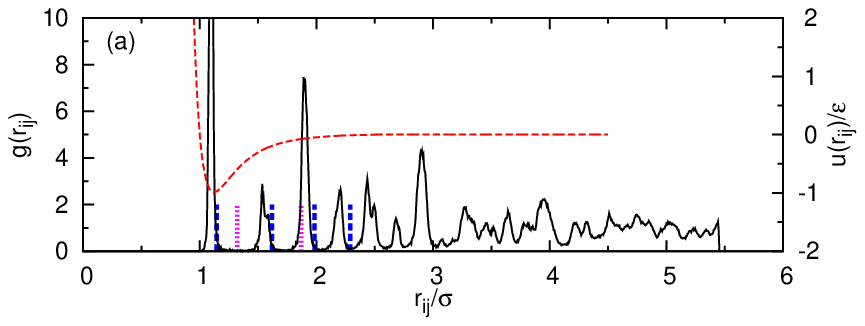}
\includegraphics[width=3.5in]{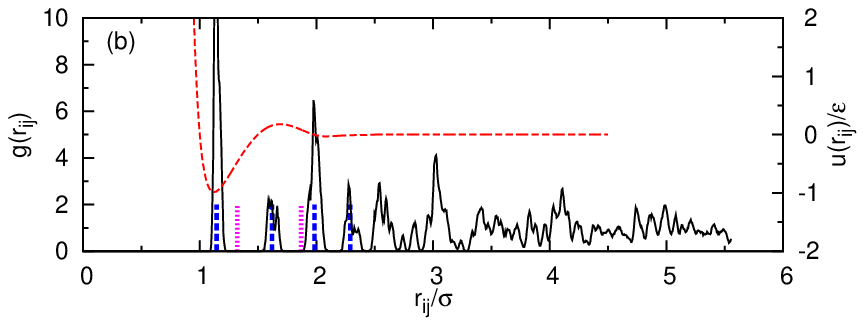}
\includegraphics[width=3.5in]{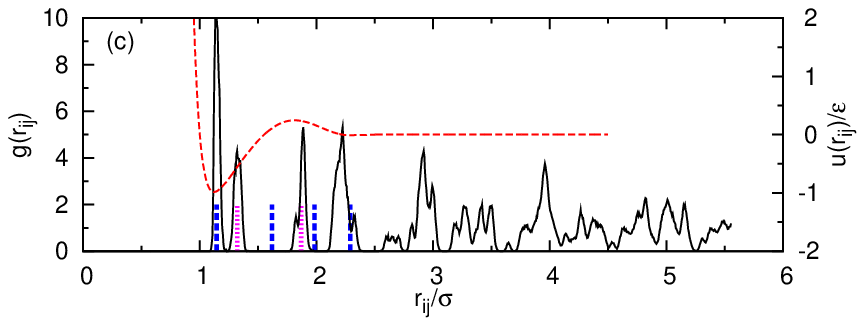}
\includegraphics[width=3.5in]{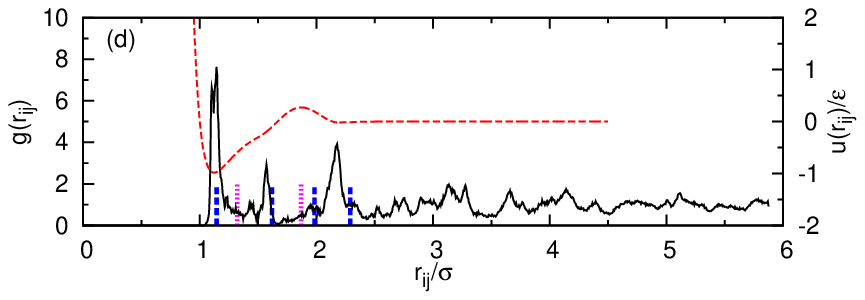}
\caption{Radial distribution function $g(r_{ij})$ (solid lines, left
axis) and pair potential $u(r_{ij})$ (dashed lines, right axis) for
monodisperse spheres that interact via (a) the Lennard-Jones potential
in a FCC crystal structure and via the DZ potential with
(b) $\lambda = \sqrt[6]{2}$ and $\delta = 2.1$ in
a FCC crystal structure, (c) $\lambda = 1.2$ and $\delta = 2.3$ in
a BCC crystal structure, and (d) $\lambda = 1.5$ and $\delta = 2.2$
in a $8$-$4$ crystal structure. The vertical dotted and
dashed lines indicate the BCC lattice spacings (relative to $r_m$)
$1$:$2/\sqrt{3}$:$2$ and the FCC lattice spacings 
$1$:$\sqrt{2}$:$\sqrt{3}$:$2$ up to third and fourth nearest neighbors, 
respectively.}
\label{fig:grDZ}
\end{figure}

We also investigated crystallization and glass formation as a function
of the form of intermediate-range repulsive pairwise interactions
(Sec.~\ref{dz_section}).  We first performed molecular dynamics simulations
of monodisperse spheres interacting via the DZ potential
(Eq.~\ref{eq:uds}) at fixed strength $\xi=0.35\epsilon$ and varying peak
location $(\lambda+\delta)/2$ and width $\delta-\lambda$. In
Fig.~\ref{fig:RcDZ}, we plot the critical cooling rate $R_c$ as a
contour plot versus $(\lambda+\delta)/2$ and $\delta-\lambda$ over
ranges that are relevant to BMGs (Table~\ref{table:bump}).  We find
several regions of good glass-forming ability (small $R_c$) and
different crystal structures that compete with glass formation.  For a
large region of parameter space, FCC is the competing crystal
structure.  BCC is the competing crystal structure when the location
of the peak in $u_{\rm bump}$ approaches third-neighbor separations at
$r_{ij} \approx \sqrt{3} r_m$. We also find an ``8-4'' crystal
structure that competes with glass formation, with atom positions
located on embedded octagons and squares when they are projected into
two dimensions. (See the inset of Fig.~\ref{fig:RcDZ}). In three
dimensions, one can see that the atoms forming the octagons and
squares are located in alternating stacked layers.  (See Fig.~\ref{fig:grDZ}
for a comparison of the radial distribution functions for FCC, BCC, and 
$8$-$4$ crystals.) When the
intermediate-range repulsion becomes too strong ({\it i.e.} large $\delta$), microphase separation becomes energetically favorable
compared to macroscale phase
separation~\cite{brazovskii:1975,seul:1995}.

We also studied the critical cooling rate $R_c$ for binary mixtures
({\it e.g.} Zr-Cu alloys), in which one component possesses
intermediate-range repulsive interactions and the other component does not.
We focused on binary systems with atoms that interact via the DZ ($A$
species) and LJ potential ($B$ species) with diameter ratio
$\sigma_B/\sigma_A=0.95$.  For the DZ potential, we set the parameters
$\xi /\epsilon \approx 0.4$, $\lambda \approx 1.2$, and $\delta \approx 2.2$ to
mimic those of Zr atoms (Table~\ref{table:bump}). As shown in
Fig.~\ref{fig:rcbinary}, $R_c$ for this binary mixture is suppressed
by more than two orders of magnitude compared to the pure system with
LJ or DZ interactions alone because the two species possess
incompatible equilibrium crystal structures ({\it i.e.}  FCC and
BCC).  This mechanism of incompatible equilibrium crystal structures
may explain the exceptionally good glass-forming ability of the Zr-Cu
system, even though it is a binary, rather than, multi-component alloy.

\subsection{LJ EAM for Monoatomic Systems}
\label{mbe}

\begin{figure}
\includegraphics[width=3.5in]{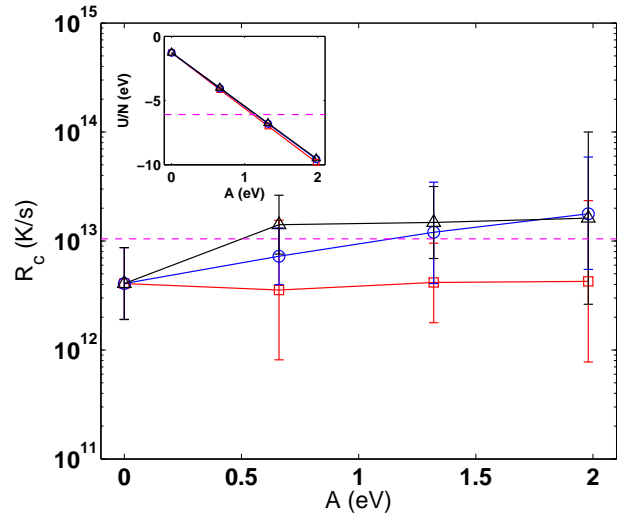}
\caption{The critical cooling rate $R_c$ from simulations of the LJ-EAM
plotted as a function of the many-body interaction strength $A$ (in
eV) for several values of the electron density inverse decay length
$\beta=2$ (squares), $4$ (circles), and $6$~${\AA}^{-1}$
(triangles). $R_c$ from simulations of the full EAM for Zr ({\it i.e.} 
$A \approx 1.32$ eV and 
$\beta \approx 4$~${\AA}^{-1}$) is indicated by the horizontal dashed
line. Error bars give the standard deviation from $10$ independent simulations with random 
initial conditions. The inset shows the total potential energy $U/N$ per atom 
(for cooling rates $R > R_c$) versus $A$ 
for the same values of $\beta$ in the main panel. $U/N$ for the full EAM 
of Zr is given by the horizontal dashed line.}
\label{fig:Rc_LJEAM}
\end{figure}

To determine the relative contributions of the pairwise and many-body
interactions to the GFA of alloys, we performed molecular dynamics
simulations of the LJ-EAM potential (Sec.~\ref{eam_section}) as a
function of the many-body interaction strength $A$ and electron
density inverse decay length $\beta$ for monoatomic systems. In
Fig.~\ref{fig:Rc_LJEAM}, we show the critical cooling rate $R_c$ for
monodisperse LJ-EAM systems as a function of $A$ for $\beta = 2$, $4$,
and $6$~${\AA}^{-1}$. We find that $R_c \approx 10^{13}~{\rm K/s}$.  $R_c$
changes by less than one order of magnitude as $A$ and $\beta$ are
varied over the range that is relevant for elements found in BMGs even
though the total potential energy per atom $U/N$ varies linearly with
$A$. We also find that FCC crystals are the ordered structures that
compete with glass formation in monoatomic LJ-EAM systems over the
full parameter range for $A$ and $\beta$. Thus, we argue that
many-body interactions have a weak influence on the GFA compared to
the pairwise interactions for monoatomic systems.

\subsection{Full EAM for Binary Alloys}
\label{full_eam}

\begin{figure}
\includegraphics[width=3.7in]{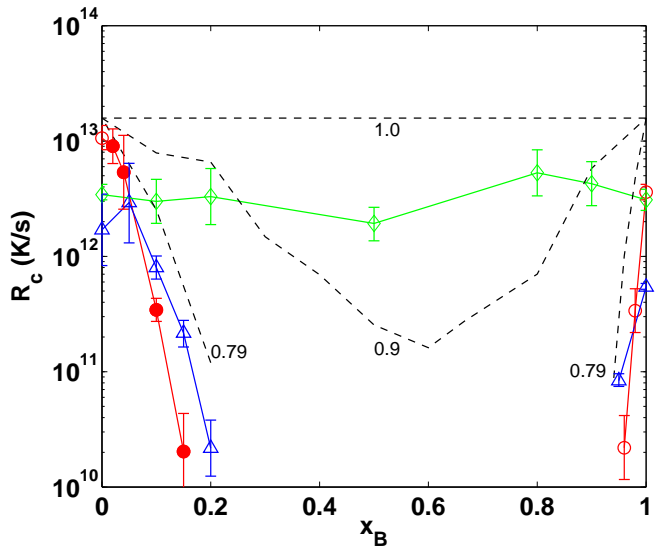}
\caption{The critical cooling rate $R_c$ (in K/s) for several binary
alloys, including Zr-Cu with atomic diameter ratio $\alpha=0.79$
(circles), Mg-Al with $\alpha = 0.94$ (triangles), and Cu-Ni with
$\alpha=0.98$ (diamonds), is plotted as a function of number
fraction of small atoms $x_B$ using molecular dynamics simulations
of the full EAM. Error bars on $R_c$ are obtained from the standard
deviation from $5$ independent simulations. The EAM source files are
given in
Refs.~\cite{sheng:2011,foiles:1985,zhou:2004,liu:1998,mendelev:2009}. As
a comparison, $R_c$ for additive binary hard spheres with $\alpha =
1.0$, $0.9$, and $0.79$ are shown as dashed lines. We also
indicate when FCC or HCP (open symbols) and BCC (filled symbols)
crystal structures compete with glass formation.}
\label{fig:Rc_EAM}
\end{figure}

We also measured the critical cooling rate $R_c$ for several binary
alloys as a function of the number fraction $x_B$ of the small atomic
species using the full EAM potential. We focused on Zr-Cu, Mg-Al,
and Cu-Ni alloys with atomic diameter ratios that range from
$\alpha = 0.79$ to $0.98$.  In Fig.~\ref{fig:Rc_EAM}, we compare $R_c$
versus $x_B$ from simulations of the full EAM potential for these
alloys to $R_c$ obtained from simulations of additive hard spheres with
comparable values of $\alpha$~\cite{zhang:2014}.

As expected, $R_c$ for binary alloys with $\alpha \sim 1$ ({\it i.e.}
Cu-Ni) is nearly independent of $x_B$.  In addition, when the
hard-sphere simulations with $\alpha = 1$ are calibrated to Ni, $R_c$
from simulations of the hard-sphere and EAM potentials agree
semi-quantitatively.  From our previous simulations of hard
spheres~\cite{zhang:2014}, we know that $R_c(x_B)$ develops a deep
minimum that shifts to larger $x_B$ as $\alpha$ decreases from unity.
For example, when $\alpha =0.9$, $R_c$ for hard-sphere systems at $x_B
\approx 0.6$ is two orders of magnitude less than the value when
$\alpha=1$.  Although we are not able to simulate sufficiently slow
rates, it appears that $R_c$ at the minimum in $x_B$ for Mg-Al with
$\alpha=0.94$ will decrease by at least two orders of magnitude and
the minimum in $R_c(x_B)$ will occur at $x_B > 0.5$.  We also 
find similar results for $R_c$ for hard spheres with $\alpha = 0.79$ 
and for EAM of Zr-Cu with a deep minimum in the range $0.2 < x_B < 0.8$. 

We also determined the crystal structures that compete with glass
formation in the full EAM simulations of binary alloys. We find that
FCC (or HCP) is most often the competing crystal structure, as in
simulations of additive binary hard spheres, but we also find
exceptions. In particular, we show that on the Zr-rich side of Zr-Cu,
BCC crystal structures compete with glass formation.  The BCC
equilibrium structure for the Zr-Cu alloys can likely be attributed to
the pairwise part of the EAM potential.  For example, the pair
potential for Zr possesses intermediate-range repulsive interactions with
the location of peak $(\lambda+\delta)/2=1.70$ and width
$\delta-\lambda=1.08$ (Table~\ref{table:bump}) in a region of
parameter space that has been shown to display BCC crystal structure
(Fig.~\ref{fig:RcDZ}). 

\section{Conclusion}
\label{sec:conclusion}

The hard-sphere model has provided a predictive description of
crystallization and glass formation in simple
liquids~\cite{weeks:1971}.  In addition, we have shown in recent
studies that the additive hard-sphere model can explain more than $13$
orders of magnitude variation in the critical cooling rate $R_c$,
which nearly spans the full range of GFA from that for pure metals to that 
for the best
BMGs~\cite{zhang:2014}. We also showed that the best binary and
ternary BMGs occur in the region of parameter space ({\it i.e.}
diameter ratio and number fraction) with the smallest values of $R_c$
for hard spheres.

However, in metallic systems, there are a number of additional
features of the interatomic potential beyond hard-core repulsions,
including softness, non-additivity, and range of the pairwise
interactions.  For example, metallic atoms typically
appear softer (with smaller values of the exponent of the repulsive
core) than the commonly used LJ pair potential and possess several per
cent negative non-additivity due to shortening of metallic
bonds~\cite{cheng:2009}. In addition, Friedel oscillations in metals 
give rise to intermediate-range repulsion at separations beyond the short-range
attractive well~\cite{Friedel:1958}. The interatomic potential for metals 
also includes many-body interactions from the electronic degrees of 
freedom. In this manuscript, we investigated how these additional
features affect the GFA of pure and binary metallic
systems.

We performed molecular dynamics simulations of several model systems
to study the effects on the GFA for each of the key features of the
interatomic potential separately.  For example, we performed
simulations of monodisperse and binary spheres that interact via the
generalized LJ and DZ pair potentials to quantify the effect of the softness
of the repulsive core and form of the intermediate-range repulsive interactions on the
GFA.  We also performed MD simulations of non-additive binary hard
spheres to quantify the effects of non-additivity on the GFA. We
found that softness, non-additivity, and form of the
intermediate-range repulsions cause deviations in $R_c$ that are only
$1\sim 2$ orders of magnitude from the additive hard-sphere
predictions.

While FCC is the most stable crystal structure for LJ and hard-sphere
systems, softening of the repulsive core gives rise to novel contracted
disordered structures, as well as the formation of BCC  
crystals. We also showed that negative non-additivity of the
repulsive core in binary alloys improves the GFA when the competing
crystal structures are solid solutions. However, when the atomic size
ratio is in the demixing regime ($\alpha < 0.8$), negative
non-additivity can favor the formation of compound crystals and
decrease the GFA. The crystal structure that competes with glass
formation, and thus the GFA, also depends sensitively on the form
of the intermediate-range repulsive interactions. We find that when the
competing crystal structures of each component in an alloy are
incompatible ({\it e.g.} FCC and BCC), the GFA can be enhanced
compared to hard-sphere predictions.

We also investigated the relative contributions of the pairwise and
many-body interactions to the GFA by performing molecular dynamics
simulations of the LJ-EAM potential. We found that including the
many-body interactions only changes $R_c$ by less than one order of
magnitude compared to that when the many-body interactions are not
included.  We also calculated $R_c$ for several binary alloys using
the full EAM potential and found qualitatively the same results as for
binary hard spheres. Thus, we argue that hard-sphere interactions
provide a qualitatively accurate model for predicting the GFA of
alloys. Other features of the interatomic potential (beyond 
additive hard-core repulsion) give rise to only
$1$-$2$ orders of magnitude variation of $R_c$, which is small
compared to the more than $13$ orders of magnitude variation predicted
by hard-sphere systems.  Despite this, including additional features to the
interatomic potential beyond hard-sphere interactions is important for
the design of new BMGs since precise quantification of the critical casting 
thickness can determine whether a new BMG is commercially viable.

\begin{acknowledgments}
The authors acknowledge primary financial support from the NSF MRSEC
DMR-1119826 (KZ) and partial support from NSF grant numbers
CMMI-1462439 (CO, MF) and CMMI-1463455 (MS). This work also benefited from
the facilities and staff of the Yale University Faculty of Arts and
Sciences High Performance Computing Center and the NSF (Grant
No. CNS-0821132) that in part funded acquisition of the computational
facilities.
\end{acknowledgments}


\end{document}